# The transport of relative canonical helicity

S. You

*Aeronautics & Astronautics, University of Washington, Seattle, WA 98195, USA*

The evolution of relative canonical helicity is examined in the two-fluid magnetohydrodynamic formalism. Canonical helicity is defined here as the helicity of the plasma species' canonical momentum. The species' canonical helicity are coupled together and can be converted from one into the other while the total gauge-invariant relative canonical helicity remains globally invariant. The conversion is driven by enthalpy differences at a surface common to ion and electron canonical flux tubes. The model provides an explanation for why the threshold for bifurcation in counter-helicity merging depends on the size parameter. The size parameter determines whether magnetic helicity annihilation channels enthalpy into the magnetic flux tube or into the vorticity flow tube components of the canonical flux tube. The transport of relative canonical helicity constrains the interaction between plasma flows and magnetic fields, and provides a more general framework for driving flows and currents from enthalpy or inductive boundary conditions.

## I. INTRODUCTION

Previous treatments of canonical helicity—also known as generalized vorticity [1], self-helicity [2], generalized helicity [3], or fluid helicity [4]—concluded that the canonical helicities of each species were invariant, independent from each other. Assuming closed canonical circulation flux tubes inside singly-connected volumes, and arguing for selective decay arguments in the presence of dissipation, it was shown that canonical helicity is a constant of the system stronger than magnetofluid energy. Generalized relaxation theories could therefore minimize magnetofluid energy for a given canonical helicity and derive stationary relaxed states.

It is puzzling that in a multiple-component plasma a species' canonical helicity must be independent from another. Given a general isolated system, the canonical momentum vectors of ion and electron fluid elements trace out different closed helical paths. Both paths are linked, resembling intertwined helical braids directed along a magnetic field line, and define flux tubes of canonical momentum that interpenetrate each other. On scales larger than the ion and electron skin depths, defined as $c/\omega_{p\sigma}$ where $c$ is the speed of light and $\omega_{p\sigma}$ is the plasma frequency of the species $\sigma$, or when species momentum is negligible, canonical flux tubes are topologically indistinguishable from magnetic flux tubes, so magnetic helicity suffices to describe the quasi-static evolution of the system. But when species momentum is significant or when phenomena involves scales that include ion or electron skin depths, then canonical flux tubes are distinguishable from one another, and because they overlap, any effort to count helicity in the system should consider gauge dependence.

This paper therefore examines the evolution of gauge invariant relative helicity (Sections II and III) and argues that each species' helicities are linked, even in isolated singly-connected volumes (Section IV). A decrease in one species' canonical helicity is compensated by an increase in the other species' canonical helicity. The transfer of canonical helicity explains why the experimental eigenvalue threshold depends on the size parameter during the observed bifurcation of the final compact torus configuration formed from counter-helicity merging (Section V). In addition to coupling different species, canonical helicity can be explicitly expressed as the weighted sum of magnetic, kinetic and cross-helicities. The evolution equations (Section VI) show that magnetic helicity can be converted into helical flows, and vice versa, which can be interpreted as reconnection of magnetic flux tubes to vorticity flow tubes that preserves the total canonical helicity of the system.

## II. RELATIVE CANONICAL HELICITY

By direct analogy with relative magnetic helicity [5, 6], a canonical helicity has topological meaning and is gauge invariant only if all the flux tubes of canonical vorticity $\vec{\Omega}_\sigma = \nabla \times \vec{P}_\sigma$ are completely enclosed inside the closed volume $V$ under consideration, where $\vec{P}_\sigma = m_\sigma \vec{u}_\sigma + q_\sigma \vec{A}$ is the canonical momentum of a fluid element of species $\sigma$ with mass $m_\sigma$, charge $q_\sigma$, flowing at velocity $\vec{u}_\sigma$ in a magnetic field with vector potential $\vec{A}$. The fluid vorticity is defined as $\vec{\omega}_\sigma = \nabla \times \vec{u}_\sigma$ and the magnetic field $\vec{B} = \nabla \times \vec{A}$. If any part of a flux tube penetrates the surface $S$ bounding the volume $V$, a relative canonical helicity should be used in order to remove gauge ambiguity. The gauge invariant relative canonical helicity can be defined in similar fashion to relative magnetic helicity [5] as

$$K_{\sigma rel} = \int_V \vec{P}_{\sigma-} \cdot \vec{\Omega}_{\sigma+} \, dV \quad (1)$$

where the plus and minus subscripts refer to the combination $\vec{X}_\pm = \vec{X} \pm \vec{X}_{ref}$ of some actual field $\vec{X}$ with a chosen reference vector field $\vec{X}_{ref}$. For example, the reference canonical momentum is

$$\vec{P}_{\sigma ref} = m_\sigma \vec{u}_{\sigma ref} + q_\sigma \vec{A}_{ref} \quad (2)$$

such that $\vec{P}_{\sigma\pm} = m_\sigma \vec{u}_{\sigma\pm} + q_\sigma \vec{A}_\pm = \vec{P}_\sigma \pm \vec{P}_{\sigma ref}$. The physical properties are fixed to the nature of the actual species not the reference species, i.e. $m_\sigma$ does not become $m_{\sigma ref}$ and $q_\sigma$ does not become $q_{\sigma ref}$. Scalar fields $\phi$ are combined in a similar manner with a reference field $\phi_{ref}$. To remove gauge dependence, the arbitrary reference fields are chosen to have the same normal components on the surface $S$ as the actual fields, so $\vec{X}_{ref} \cdot d\vec{S} = \vec{X} \cdot d\vec{S}$ and $\nabla \phi \cdot d\vec{S} = \nabla \phi_{ref} \cdot d\vec{S}$ on the surface.

The canonical helicity can be expanded to

$$K_{\sigma rel} = m_\sigma^2 \mathcal{H}_{\sigma rel} + m_\sigma q_\sigma \mathcal{X}_{\sigma rel} + q_\sigma^2 \mathcal{K}_{rel} \quad (3)$$

which is a weighted sum of relative magnetic helicity $\mathcal{K}_{rel} = \int_V \vec{A}_- \cdot \vec{B}_+ \, dV$, relative kinetic helicity $\mathcal{H}_{\sigma rel} = \int_V \vec{u}_{\sigma-} \cdot \vec{\omega}_{\sigma+} \, dV$, and relative cross-helicity, $\mathcal{X}_{\sigma rel} = \int_V \left( \vec{u}_{\sigma-} \cdot \vec{B}_+ + \vec{u}_{\sigma+} \cdot \vec{B}_- \right) dV$. The relative magnetic helicity $\mathcal{K}_{rel}$ depends on the chosen electromagnetic reference field but does not depend on species differentiation, so the subscript $\sigma$ can be dropped and $\vec{A}_{ref}$ is the same for both species (e.g. a vacuum field $\vec{A}_{vac}$). The magnetic helicity contribution to each species' canonical helicity is independent of the sign of the charge and equally weighted if the plasma is singly-ionized. The relative kinetic helicity $\mathcal{H}_{\sigma rel}$ depends on an explicit differentiation of the two species but not on the electromagnetic reference field. The flow field of one species can therefore be chosen as a reference for the other species, say $\vec{u}_{\sigma ref} \rightarrow \vec{u}_\alpha$ for a two-component plasma $\sigma$ and $\alpha$, which results in equal and opposite kinetic helicities $\mathcal{H}_{\sigma rel} = -\mathcal{H}_{\alpha rel}$ at all times. This is an expression of the fact that ion and electron flows are intertwined and topologically mirrored. The contribution of kinetic helicities to the species canonical helicity depends quadratically on the mass $m_\sigma$ and thus favours ion canonical helicity. The cross-helicity $\mathcal{X}_{\sigma rel}$ depends on species differentiation and the electromagnetic reference field. Since the reference electromagnetic field $\vec{A}_{ref}$ can be chosen such that $\vec{A}_- \times d\vec{S} = 0$ on the surface boundary $S$, the cross-helicity can thus be written $\mathcal{X}_{\sigma rel} = 2 \int_V \left( \vec{u}_\sigma \cdot \vec{B} - \vec{u}_{\sigma ref} \cdot \vec{B}_{ref} \right) dV = \mathcal{X}_\sigma - \mathcal{X}_{\sigma ref}$, where $\mathcal{X}_\sigma = 2 \int_V \vec{u}_\sigma \cdot \vec{B} \, dV$ is the ordinary cross-helicity for closed systems ($\vec{A}_{ref} = 0$). Finally, a total relative canonical helicity is defined as $\mathbb{K}_{rel} = \sum_\sigma K_{\sigma rel}$.

### III. EVOLUTION OF RELATIVE CANONICAL HELICITY

The relative canonical helicity transport equations are based on the two-fluid equations of motion

$$n_\sigma m_\sigma \frac{d\vec{u}_\sigma}{dt} = n_\sigma q_\sigma \left( \vec{E} + \vec{u}_\sigma \times \vec{B} \right) - \nabla \mathcal{P}_\sigma - \vec{R}_\sigma \tag{4}$$

where the general frictional force $\vec{R}_\sigma = \vec{R}_{\sigma\alpha} + \vec{R}_{\sigma\sigma} + \vec{R}_{th\sigma}$ includes the interspecies drag force $\vec{R}_{\sigma\alpha} \equiv \nu_{\sigma\alpha} n_\sigma m_\sigma (\vec{u}_\sigma - \vec{u}_\alpha)$ due to collisions at frequency $\nu_{\sigma\alpha}$ between species $\sigma$ and $\alpha$, a viscous contribution $\vec{R}_{\sigma\sigma} \equiv \nabla \cdot \left( \overleftrightarrow{\Pi}_\sigma - \mathcal{P}_\sigma \hat{I} \right)$, and any thermal Nernst effect contribution $\vec{R}_{th\sigma}$. An isolated two-fluid plasma consisting only of ions $i$ and electrons $e$ cannot change its own momentum, so $\vec{R}_{ei} + \vec{R}_{ie} = 0$. The scalar isotropic pressure $\mathcal{P}_\sigma$ is isolated from the pressure tensor $\overleftrightarrow{\Pi}_\sigma$ to define the generalized enthalpy as $h_\sigma \equiv q_\sigma \phi + 1/2 \, m_\sigma u_\sigma^2 + \int d\mathcal{P}_\sigma / n_\sigma$, combining the work contribution from an electrostatic potential $\phi$ and the mechanical enthalpy for a barotropic plasma where the species' pressure is a function of the species' density only, $\mathcal{P}_\sigma = \mathcal{P}_\sigma(n_\sigma)$. In canonical form, the equations of motion become

$$\frac{\partial \vec{P}_\sigma}{\partial t} + \vec{\Omega}_\sigma \times \vec{u}_\sigma = -\nabla h_\sigma - \frac{\vec{R}_\sigma}{n_\sigma} \tag{5}$$

and the circulation of Eq. (5) gives the canonical induction equation

$$\frac{\partial \vec{\Omega}_\sigma}{\partial t} + \nabla \times (\vec{\Omega}_\sigma \times \vec{u}_\sigma) = -\nabla \times \left(\frac{\vec{R}_\sigma}{n_\sigma}\right) \tag{6}$$

which shows that a species' canonical vorticity is frozen to its fluid motion in the absence of frictional circulation. If viscosity and thermal effects can be neglected $\vec{R}_{\sigma\sigma} \sim 0, \vec{R}_{th\sigma} \sim 0$ but interspecies friction cannot be neglected so $\vec{R}_\sigma \simeq \vec{R}_{\sigma\alpha}$, a species' canonical vorticity becomes (collisionally) coupled to the other species' canonical vorticity, and because the plasma cannot change its own momentum, Eq. (6) becomes

$$n_\sigma \left(\frac{\partial \vec{\Omega}_\sigma}{\partial t} + \nabla \times [\vec{\Omega}_\sigma \times \vec{u}_\sigma]\right) = -n_\alpha \left(\frac{\partial \vec{\Omega}_\alpha}{\partial t} + \nabla \times [\vec{\Omega}_\alpha \times \vec{u}_\alpha]\right) \tag{7}$$

In the presence of interspecies friction, canonical vorticity is not frozen to the species' fluid but will decrease or increase in the opposite manner to the change in canonical vorticity of the other fluid. Eq. (7) also shows that the rate of change of canonical vorticity into another canonical vorticity can be varied by changing the charge neutrality ratio, $n_e/n_i$ in an ion-electron plasma.

A canonical electric field $\vec{\mathbb{E}}_\sigma \equiv -\nabla h_\sigma - \partial \vec{P}_\sigma/\partial t$ can be built from an enthalpy gradient and a canonical inductive term. The canonical electric field plays the role of the usual electric field $\vec{E} = -\nabla \phi - \partial \vec{A}/\partial t$ up to a constant in the canonical equation of motion, so Eq. (5) becomes an Ohm's law for canonical quantities

$$\vec{\mathbb{E}}_\sigma + \vec{u}_\sigma \times \vec{\Omega}_\sigma = \frac{\vec{R}_\sigma}{n_\sigma} \tag{8}$$

and the canonical induction equation Eq. (6) simplifies to

$$\frac{\partial \vec{\Omega}_\sigma}{\partial t} = -\nabla \times \vec{\mathbb{E}}_\sigma \tag{9}$$

which is Faraday's law for canonical quantities.

Because enthalpy is naturally useful as a relative quantity $h_{\sigma-} \equiv h_\sigma - h_{\sigma ref} = q_\sigma \phi_- + 1/2\, m_\sigma (u_\sigma^2)_- + (\int d\mathcal{P}_\sigma/n_\sigma)_-$, relative canonical helicity is defined with $\vec{P}_{\sigma-}$ as in Eq. (1) instead of opposite subscripts $\vec{P}_{\sigma+} \cdot \vec{\Omega}_{\sigma-}$ as defined for relative magnetic helicity. Both formulations are gauge invariant. Adding and subtracting actual and reference versions of Eq.(4), expanding the total derivative into partial and convective terms and using the definition of the electric field gives the relative versions of Eq. (5),

$$\frac{\partial \vec{P}_{\sigma\pm}}{\partial t} - (\vec{u}_\sigma \times \vec{\Omega}_\sigma \pm \vec{u}_{\sigma ref} \times \vec{\Omega}_{\sigma ref}) = -\nabla h_{\sigma\pm} - \frac{\vec{R}_{\sigma\pm}}{n_\sigma}. \tag{10}$$

where the reference enthalpy is defined as $h_{\sigma ref} = q_\sigma \phi_{ref} + 1/2\, m_\sigma u_{\sigma ref}^2 + \int d\mathcal{P}_\sigma/n_\sigma$, consistent with thermodynamic definitions of enthalpy. This reference enthalpy is obtained from the reference version of Eq. (4) for a relative flow field $n_\sigma m_\sigma \vec{u}_{\sigma ref}$ with a reference Lorentz force $q_\sigma(\vec{E}_{ref} + \vec{u}_{\sigma ref} \times \vec{B}_{ref})$ and reference $\nabla \mathcal{P}_{\sigma ref}$, $\vec{R}_{\sigma ref}$ forces. In principle, as long as boundary conditions for gauge invariance are satisfied and the gauge is analytic within the region of interest, one can choose any definition of a reference vector field for canonical momentum because helicity has a topological nature rather than a physical (metric) nature, i.e. the momentum field $m_\sigma \vec{u}_{\sigma ref}$ is topologically but not metrically equivalent to the flow field $\vec{u}_{\sigma ref}$ [7]. This distinction is not necessary in relative magnetic helicity transport because the magnetic vector potential is evolved without any multiplying physical constants, nor was it necessary in self-helicity concepts [1, 2] because relativeness was not considered. However, for the evolution of relative canonical helicity, the choice of reference field for the canonical momentum *does* have an effect on the definition of enthalpy and thus on the evolution of $\vec{P}_{\sigma\pm}$ (a metric quantity), even as $K_{\sigma rel}$ (a topological quantity) remains the same. For example, if $\vec{P}_{\sigma ref} \equiv m_{\sigma ref} \vec{u}_{\sigma ref} + q_{\sigma ref} \vec{A}_{ref}$ is chosen instead of Eq. (2), $\vec{P}_{\sigma-}$ evolves with a similar equation to Eq. (10) but with an enthalpy gradient defined instead with $h_{\sigma-} = q_\sigma \phi_- + 1/2 \, m_\sigma (\vec{u}_{\sigma-})^2 + (\int d\mathcal{P}_\sigma / n_\sigma)_-$. In addition to being inconsistent with the thermodynamic definition of enthalpy, this formulation implies that the total canonical helicity $\mathbb{K}_{rel}$ is zero simply by virtue of Eq. (1), when the canonical momentum of species $\alpha$ is taken as reference for the other species $\sigma$, $\vec{P}_{\sigma ref} \to \vec{P}_\alpha$ (see Section IV). The definition in Eq. (2) is thermo-dynamically valid and rests on only two reference fields, the reference velocity field $\vec{u}_{\sigma ref}$ and the reference magnetic vector potential $\vec{A}_{ref}$, without referring to a further number of reference scalar fields $m_{\sigma ref}$, $q_{\sigma ref}$ or $n_{\sigma ref}$. Thus the choice of a reference for relative canonical helicity is fully specified with just three reference fields: a flow $\vec{u}_{\sigma ref}$, a magnetic vector potential $\vec{A}_{ref}$ and an electrostatic potential $\phi_{ref}$. The chosen $\vec{u}_{\sigma ref}$ then automatically specifies the reference frictional force $\vec{R}_{\sigma\alpha ref}$, and after extending the argument to include a random velocity component, specifies the reference pressure $\nabla \mathcal{P}_{\sigma ref}$ and viscous force $\vec{R}_{\sigma\sigma ref}$.

The time derivative of the relative canonical helicity Eq. (1) is

$$\frac{dK_{\sigma rel}}{dt} = \int_V \frac{\partial (\vec{P}_{\sigma-} \cdot \vec{\Omega}_{\sigma+})}{\partial t} dV + \int_S (\vec{P}_{\sigma-} \cdot \vec{\Omega}_{\sigma+}) \vec{u}_\sigma \cdot d\vec{S} \tag{11}$$

where the motion of the boundary $\vec{u}_\sigma$ is included in the second term on the right hand side using Leibniz' rule. Substituting for $\partial \vec{P}_{\sigma-}/\partial t = \mathbb{E}_{\sigma-} + \nabla h_{\sigma-}$ and $\partial \vec{\Omega}_{\sigma+}/\partial t = -\nabla \times \mathbb{E}_{\sigma+}$ with Eq. (9) and the definition of the canonical electric field into Eq. (11) gives

$$\frac{dK_{\sigma rel}}{dt} = -\int_V \left( \mathbb{E}_{\sigma+} \cdot \vec{\Omega}_{\sigma-} + \mathbb{E}_{\sigma-} \cdot \vec{\Omega}_{\sigma+} \right) dV - \int_S h_{\sigma-} \vec{\Omega}_{\sigma+} \cdot d\vec{S} \\ - \int_S \vec{P}_{\sigma-} \times \frac{\partial \vec{P}_{\sigma+}}{\partial t} \cdot d\vec{S} + \int_S (\vec{P}_{\sigma-} \cdot \vec{\Omega}_{\sigma+}) \vec{u}_\sigma \cdot d\vec{S} \tag{12}$$

after using the solenoidal property of the canonical vorticity. The third term on the right-hand side represents a time-dependent canonical helicity injection through the surface boundary, e.g. current drive by radio-frequency (RF) waves [4]. The term can be eliminated if the boundary is a canonical flux conserver, when all normal components are constant in time on the surface of the volume. The volume integral of Eq. (12) can be expanded to $2 \int (\vec{\mathbb{E}}_{\sigma ref} \cdot \vec{\Omega}_{\sigma ref} - \vec{\mathbb{E}}_\sigma \cdot \vec{\Omega}_\sigma) dV$ and represents a generalization of the usual $2 \int \vec{E} \cdot \vec{B} \, dV$ source and sink terms in magnetic helicity transport. The canonical vorticity $\vec{\Omega}_\sigma$ plays the same role in two-fluid plasma dynamics as the magnetic field $\vec{B}$ does in center-of-mass magnetohydrodynamics (MHD). Using Eq. (8) and explicitly writing out the surface integral, Eq. (12) becomes

$$\frac{dK_{\sigma rel}}{dt} = 2 \int_V \left( \frac{\vec{R}_{\sigma ref}}{n_\sigma} \cdot \vec{\Omega}_{\sigma ref} - \frac{\vec{R}_\sigma}{n_\sigma} \cdot \vec{\Omega}_\sigma \right) dV - 2 \int_S (h_\sigma - h_{\sigma ref}) \vec{\Omega}_\sigma \cdot d\vec{S} \\ - \int_S \vec{P}_{\sigma-} \times \frac{\partial \vec{P}_{\sigma+}}{\partial t} \cdot d\vec{S} + \int_S (\vec{P}_{\sigma-} \cdot \vec{\Omega}_{\sigma+}) \vec{u}_\sigma \cdot d\vec{S} \qquad (13)$$

The frictional term (first integral on the right hand side) extends the equivalent Eq. 39 in Ref. 8 to relative canonical helicity. The enthalpy term (second term on the right-hand side) represents the generalization of magnetic helicity injection by an electrostatic potential difference across the ends of a magnetic flux tube $\dot{\mathcal{K}} \sim \int \phi \vec{B} \cdot d\vec{S}$ to the injection of canonical helicity by an enthalpy difference across the ends of a canonical vorticity flux tube $\dot{K}_\sigma \sim \int h_\sigma \vec{\Omega} \cdot d\vec{S}$. These terms are the extension of the first term on the right-hand side of Eq. 21 in Ref. 3 to relative helicity, without the assumption of zero normal boundary conditions, and the Leibniz term in Eq. (13) here is equivalent to the remainder of their right-hand side terms. The AC injection term (third term on the right-hand side involving $\vec{P}_{\sigma-} \times \partial \vec{P}_{\sigma+}/\partial t$) represents the various inductive helicity injection terms, electromagnetic for $\partial \vec{A}/\partial t$ components and other forces for $\partial \vec{u}_\sigma/\partial t$ components. The term is used for RF current drive [4] and, as developed in Sec. VII, for generating flow.

**IV.  TRANSFER AND CONSERVATION OF CANONICAL HELICITY**

For an isolated plasma inside a canonical flux-conserving volume ($\partial \vec{P}_{\sigma+}/\partial t \cdot d\vec{S} = 0$) with a fixed boundary ($\vec{u}_\sigma = 0$) and no friction ($\vec{R}_\sigma = 0$), canonical flux tubes are closed and do not intercept the surface $S$, so Eq. (13) reduces to a constant, ordinary canonical helicity $K_\sigma$ inside the volume $V$ [1, 3, 8]. In the limit of negligible species momentum ($m_\sigma \to 0$), magnetic helicity $\mathcal{K} = \int \vec{A} \cdot \vec{B} \, dV$ is thus preserved [9, 10] and in the limit of negligible Lorentz forces ($q_\sigma \to 0$), kinetic helicity $\mathcal{H}_\sigma = \int \vec{u}_\sigma \cdot \vec{\omega}_\sigma \, dV$ is also preserved [11]. However in the two-fluid plasma regime, both momentum and Lorentz forces contribute to the motion of canonical flux tubes specific to each species (FIG. 1). In regions with finite momentum, an ion canonical flux tube can effectively *separate* away from the electron canonical flux tube.

The ion volume $V_i$ made up only of the ion canonical flux tube, where $\int \vec{\Omega}_i \cdot d\vec{S}$ is constant, and the electron volume $V_e$ made up only of the electron canonical flux tube, where $\int \vec{\Omega}_e \cdot d\vec{S}$ is constant, branch away from each other in the sub-region $V_b$ where ion momentum is significant, while the volumes $V_i$ and $V_e$ still coincide in the sub-region $V_a$ where ion momentum is negligible. An example would be a current-carrying magnetic flux tube at a scale larger than the ion and electron skin depth, where ion and electron canonical flux tubes coincide with magnetic flux tubes, and transitioning to scales between the ion and electron skin depth, where ion canonical flux tubes are spatially distinct from magnetic flux tubes but the electron canonical flux tubes are not. The canonical vorticities do not intercept the overall boundary $S$ but they do penetrate the intersection surface $S_{sep}$, which divides the locations where canonical flux tubes $V_e, V_i$ are distinguishable from the remaining locations where $V_e$ and $V_i$ are indistinguishable. Therefore, in order to properly count helicity, *relative* canonical helicity must be used. The surface integrals in Eq. (13) are therefore not performed over the overall boundary $S$ englobing the total system volume $V = V_e + V_i$, but over the intersecting cut $S_{sep}$. Since both canonical vorticities have the same normals $\vec{\Omega}_i \cdot d\vec{S} = \vec{\Omega}_e \cdot d\vec{S}$ on $S_{sep}$, it is appropriate to regard the flow of one species as the reference flow for the other species, and set $\vec{u}_{iref} \to \vec{u}_e$ and $\vec{u}_{eref} \to \vec{u}_i$ in Eq. (13) for the ion and electron species respectively. Since the overall system is isolated, there is no need for reference electromagnetic fields and $\phi_{ref} = 0$ and $\vec{A}_{ref} = 0$. Otherwise potential (vacuum) electromagnetic fields can be specified as usual. Defining a comparative enthalpy as $h' = h_i - h_e$, Eq. (13) becomes

$$\frac{dK_{erel}}{dt} = -\frac{dK_{irel}}{dt} = 2\int_{S_{sep}} h'\vec{\Omega} \cdot d\vec{S} \qquad (14)$$

where $\vec{\Omega} \cdot d\vec{S} = \vec{\Omega}_i \cdot d\vec{S} = \vec{\Omega}_e \cdot d\vec{S}$ at the separation surface and quasineutrality ($n_e = n_i = n$) assumed for simplicity.

Eq. (14) shows that the total canonical helicity $\mathbb{K} = \mathbb{K}_{rel} = K_{irel} + K_{erel}$ is conserved and that canonical helicity can be transferred between the two species. The mechanism for helicity transfer is the enthalpy difference over the surface $S_{sep}$ separating the canonical flux tubes. Eq. (14) is a more fundamental statement of the principle of global helicity conservation than earlier treatments , because the statement takes into account finite momentum of each species and collisionless coupling of one species canonical helicity to another. Eq. (14) reverts to $\dot{\mathcal{K}} = 0, \dot{\mathcal{H}}_\sigma = 0$ and $\dot{K}_\sigma = 0$ in the appropriate limits: negligible species momentum ($m_\sigma \to 0$), negligible Lorentz forces ($q_\sigma \to 0$), negligible canonical flux separation ($S_{sep} \to 0$) or negligible enthalpy ($h' \to 0$), respectively.

The comparative enthalpy $h'$ in Eq. (14) used a species' enthalpy as reference for the other species ($h_{\sigma ref} = h_\alpha$). This arbitrary simplification is permitted provided $\nabla h_\alpha \cdot d\vec{S} = \nabla h_\sigma \cdot d\vec{s}$ holds on the separation surface boundary. For example, in a finite length canonical tube where $\nabla h \cdot d\vec{s} \neq 0$ only on the end surfaces $S_1$ and $S_2$, with $h$ uniform on those end surfaces, $h$ will only change along the axis of the canonical flux tube and the equality holds for both species even if the value of $h_\sigma \neq h_\alpha$ on those end surfaces. In a more general case, the reference fields are $\vec{u}_{\sigma ref} = \vec{u}_\alpha$ and $\phi_{ref} = 0$ (electromagnetically isolated), the reference enthalpies are the fictitious scalars $h_{\sigma ref} = q_\sigma \phi + 1/2\, m_\sigma u_\alpha^2 + \int d\mathcal{P}_\alpha/n_\sigma$ and the surface integral in Eq. (14) includes the comparative enthalpy

$$h' = h_\sigma - h_{\sigma ref} + h_\alpha - h_{\alpha ref} = \frac{1}{2}(m_\sigma - m_\alpha)(u_\sigma^2 - u_\alpha^2) \tag{15}$$

which is zero when $m_\sigma = m_\alpha$ or $|\vec{u}_\sigma| = |\vec{u}_\alpha|$. Canonical helicity $K_\sigma$ is thus transferred from one species to another while *exactly* preserving the total canonical helicity $\mathbb{K}$, for a two-species plasmas with equal mass, irrespective of the species-specific kinetic energy, or when there is no net current through the separation surface. The former situation exists, for example, in an electron-positron plasma and the latter in plasmas with no internal currents. Conversely, the total canonical helicity of an isolated system *can* vary ($\dot{\mathbb{K}} \lessgtr 0$) if a net current is self-generated ($|\vec{u}_\sigma| \lessgtr |\vec{u}_\alpha|$) across the separation surface $S_{sep}$, for example in complex scenarios with bootstrap currents or internal reconnection events.

To summarize, even though the overall system is isolated and helicity is a global (approximate) invariant, helicity can be transferred *across* species. Because helicity is a topological quantity, the twists and writhes of one species' canonical momentum can be transferred to another species' twist and writhes. In effect, this two-fluid model is a generalization of the usual transfer of magnetic helicity between two connected magnetic flux tubes to the transfer of canonical helicity between two connected canonical flux tubes. The transfer mechanism is a generalized battery effect [12, 13] due to enthalpy potential differences on surfaces separating the two canonical volumes, resulting in the coupling of helical magnetic fields to helical plasma flows. An example of this coupling is the bifurcation observed during the formation of compact toroids from counter-helicity merging.

## V.  BIFURCATION IN COMPACT TORUS FORMATION

The model of section IV provides a simple explanation for the helicity injection threshold observed in compact torus bifurcation experiments [14]. The TS-4 experiment produces a field-reversed or spheromak configuration from counter-helicity merging of spheromaks, depending on the poloidal eigenvalue set at the beginning of the discharge. The initial poloidal eigenvalue $\lambda_o/\lambda_{Taylor}$ is measured to be proportional to $1/S^*$ (FIG. 2), where $\lambda_0 = \mu_0 \langle \delta I/\delta \psi \rangle$ is the measured poloidal eigenvalue at the end-of-merging time, normalized to the Taylor eigenvalue ($\mu_0$ is the permeability of free space, $I$ is the poloidal current and $\psi$ is the poloidal magnetic flux), and $S^* \sim L/\rho_{Li}$ is the system size parameter [15], the scale length $L$ normalized to the ion Larmor radius or skin depth. During merging, the slingshot effect [16] generates Alfvénic toroidal flow velocities. The inboard and outboard flows are oppositely directed, $u_{\theta outb} \sim +v_A$ and $u_{\theta inb} \sim -v_A$. Between the beginning of merging (when two spheromaks are still distinct but in contact) to the end of merging (when only one compact torus is observed), the slingshot toroidal flows generate a kinetic energy change of $\Delta E_{kin} \sim 1/2\, m_i(u_{\theta outb}^2 + u_{\theta inb}^2) \sim m_i v_A^2$ across the compact torus cross-section. All else being equal, the enthalpy change is then $\Delta h' \sim \Delta E_{kin}$.

Supposing the new compact torus can be represented by a single canonical flux tube, where $\Psi \equiv \int \vec{\Omega} \cdot d\vec{S}$ is constant, with uniform enthalpy over the the cross-sectional area, i.e. the ion canonical flux tube coincides with the electron canonical flux tube, then Eq. (14) becomes

$$\frac{dK_{irel}}{dt} = -\frac{dK_{erel}}{dt} = 2m_i \Delta h' \text{\textit{f}} + 2q_i \Delta h' \psi \tag{16}$$

where the canonical flux tube is explicitly distinguished into the magnetic flux tube component $\psi \equiv \int \vec{B} \cdot d\vec{S}$ and the vorticity flux (or flow) tube component $\text{\textit{f}} \equiv \int \vec{\omega}_i \cdot d\vec{S}$. Eq. (16) states that for a given enthalpy difference across the ends of a canonical flux tube $\Psi = m_i \text{\textit{f}} + q_i \psi$, ion helicity is injected into the vorticity flux tube component and the magnetic flux tube component. The ratio between the two terms on the right-hand side can be defined as a fractional canonical helicity injection threshold $\bar{K}_{thr}$ as

$$\bar{K}_{thr} \equiv \frac{|m_i \Delta h' \text{\textit{f}}|}{|q_i \Delta h' \psi|} \sim \frac{m_i u_i}{q_i LB} \sim \frac{\rho_{Li}}{L} \sim \frac{1}{S^*} \tag{17}$$

for velocities measured of the order of thermal and Alfvén velocities $u_i \sim v_{ith} \sim v_A$ over the scale length $L$. The thermal ion gyroradius is $\rho_{Li} = m_i v_{ith}/q_i |\vec{B}|$. As the size parameter increases $S^* \gg 1$, a given enthalpy imposed at the ends of a canonical flux tube will channel canonical helicity increasingly into the magnetic flux tube component, $|m_i \Delta h' \text{\textit{f}}| \ll |e \Delta h' \psi|$.

In two-fluid flowing equilibria of compact plasmas [17], a field-reversed configuration (FRC) corresponds to a minimum energy state with finite ion canonical helicity but no electron canonical helicity ($K_i \neq 0, K_e = \mathcal{K} = 0$), and a spheromak corresponds to a minimum energy state with only finite magnetic helicity. An FRC can therefore be represented by a toroidal canonical flux tube $\Psi$ with a weak magnetic flux tube component dominated by the vortex component ($m_i \text{\textit{f}} \gg q_i \psi$), so any canonical helicity injection channeled preferentially into the vortex tube will result in an FRC. A spheromak can be represented by a toroidal canonical flux tube $\Psi$ dominated by the static magnetic component with a negligible vortex component ($m_i \text{\textit{f}} \ll q_i \psi$), so any canonical helicity injection channelled preferentially into the magnetic flux tube will form a spheromak. The size parameter in Eq. (17) determines the channeling ratio and is plotted with the experimental data in FIG. 2. For a given $S^*$, if the initial poloidal eigenvalue is above the threshold given by Eq. (17), helicity injection will preferentially be channeled into the magnetic component and result in a spheromak configuration. Otherwise helicity injection will preferentially be channeled into the vortex component and result in an FRC. At high $S^*$, the threshold window ($0 < |\bar{K}_{thr}| < 1/S^*$) becomes narrower, so the poloidal eigenvalue has to be set more precisely to make an FRC from counter-helicity merging, while at lower $S^*$, the threshold window becomes larger and it becomes easier to form an FRC from counter-helicity merging. Experiments aiming to make hydrogen or deuterium FRC's from counter-helicity merging at large $S^*$ will need to consider the threshold window when fine tuning experimental parameters.

The TS-4 bifurcation experiment showed that during counter-helicity merging, the slingshot effect from reconnection of twisted magnetic fields generated a combination of magnetic activity, ion shear flows and ion heating. The magnetic activity has been described as torsional kinetic Alfvén waves representing the dynamic unwinding of the magnetic field during reconnection [18, 19]. A fraction of the magnetic energy is lost to ion Landau damping and heats the ions [18]. The remainder of the magnetic energy is transferred to flows in one of two ways. In the MHD-dominated regime (high $S^*$), the driven flows are weak and below the shear stabilization threshold [20] so a fraction of the magnetic energy remains in the magnetic activity, which stays undamped, only cascading from higher modes to lower modes. The lower modes restore the toroidal magnetic field presumably by poloidal flux amplification [21], and forms the final spheromak. In the two-fluid- or kinetic-dominated regime (low $S^*$), the driven flows are strong enough to be above the shear stabilization threshold. Magnetic energy is converted to flow energy, so magnetic activity is damped, preventing the cascade to lower modes and restoration of the toroidal magnetic field. The final configuration is then an FRC with no (or little) toroidal magnetic field. The canonical helicity injection model explains why the choice of one path or the other depends on $1/S^*$, using the ratio of helicity injection into the vorticity flux tube component over the magnetic flux tube component of the canonical flux tube.

## VI. COUPLING BETWEEN MAGNETIC, KINETIC AND CROSS-HELICITIES

The canonical helicity of each species $K_\sigma$ combines the species' kinetic helicity $\mathcal{H}_\sigma$, the cross-helicity $\mathcal{X}_\sigma$ and the magnetic helicity $\mathcal{K}$, as in Eq. (3). Adding the equation of motion Eq. (4) dotted with $\vec{\omega}_\sigma$ and the curl of Eq. (4) dotted with $\vec{u}_\sigma$, then integrating using Leibniz' rule, the evolution of species' kinetic helicity can be written

$$m_\sigma \frac{d\mathcal{H}_\sigma}{dt} = 2\int \vec{\omega}_\sigma \cdot \frac{\vec{F}_{\sigma nc}}{n_\sigma} dV - \int \vec{u}_\sigma \times \frac{\vec{F}_{\sigma nc}}{n_\sigma} \cdot d\vec{S} - \int h_\sigma \vec{\omega}_\sigma \cdot d\vec{S} + m_\sigma \int (\vec{u}_\sigma \cdot \vec{\omega}_\sigma) \vec{u}_\sigma \cdot d\vec{S} \qquad (18)$$

where the non-conservative forces have been amalgamated into $\vec{F}_{\sigma nc} = n_\sigma q_\sigma (-\partial \vec{A}/\partial t + \vec{u}_\sigma \times \vec{B}) - \vec{R}_\sigma$. Kinetic helicity is conserved only if the forces acting on the fluid are conservative and the system is closed. In a magnetized plasma, the non-conservative parts of the Lorentz force and enthalpy differences across intersecting canonical flux tubes will change kinetic helicity. If the total canonical helicity is fixed under the conditions of Eq. (14), then kinetic helicity *must* be converted into other forms of helicity. Using the appropriate vector identities and multiplying with $m_\sigma$, Eq. (18) can be expanded to

$$\begin{aligned} m_\sigma^2 \frac{d\mathcal{H}_\sigma}{dt} = & -m_\sigma q_\sigma \int \vec{u}_\sigma \times \frac{\partial \vec{A}}{\partial t} \cdot d\vec{S} + 2m_\sigma q_\sigma \int \vec{\omega}_\sigma \cdot (\vec{u}_\sigma \times \vec{B}) dV - 2m_\sigma \int \vec{\omega}_\sigma \cdot \frac{\vec{R}_\sigma}{n_\sigma} dV \\ & - 2m_\sigma q_\sigma \int \vec{u}_\sigma \cdot \frac{\partial \vec{B}}{\partial t} dV - m_\sigma^2 \int \vec{u}_\sigma \times \frac{\partial \vec{u}_\sigma}{\partial t} \cdot d\vec{S} - m_\sigma q_\sigma \int \nabla \cdot (\vec{A} \times \vec{u}_\sigma) \vec{u}_\sigma \cdot d\vec{S} \\ & - m_\sigma q_\sigma \int \vec{A} \times \frac{\partial \vec{u}_\sigma}{\partial t} \cdot d\vec{S} - 2m_\sigma \int h_\sigma \vec{\omega}_\sigma \cdot d\vec{S} + m_\sigma^2 \int (\vec{u}_\sigma \cdot \vec{\omega}_\sigma) \vec{u}_\sigma \cdot d\vec{S} \end{aligned} \qquad (19)$$

expressed in a form convenient for comparison with the expansion of the canonical helicity Eq.(13). Because of the chosen definition of cross-helicity in Section II, a term is eliminated from Eq. (19). The center-of-mass (MHD) kinetic helicity is related to the species' kinetic helicity by $\mathcal{H} = (n_i m_i \mathcal{H}_i + n_e m_e \mathcal{H}_e)/nm$, where $nm = n_i m_i + n_e m_e$ and the time evolution $\dot{\mathcal{H}}$ can be retrieved by summing Eq. (18) for each species in this way, and noting that the total derivatives are defined using a convection with each species $d/dt = \partial/\partial t + \vec{u}_\sigma \cdot \nabla$, so must be transformed into the center-of-mass frame. The final form for the center-of-mass kinetic helicity contains the usual MHD kinetic helicity transport terms $\int d(\vec{u} \cdot \vec{\omega})/dt \, dV$, where $\vec{u}$ is the center-of-mass velocity and $\vec{\omega} = \nabla \times \vec{u}$, a term for kinetic helicity injection by enthalpy through vorticity flux tube surfaces $\int [(m_i m_e/m^2 n^2 e^2) J^2/2 + u^2/2] \, \vec{\omega} \cdot d\vec{S}$, where $\vec{J}$ is the current density, and terms which include integrals involving $\nabla \times \vec{J}$. The $\nabla \times \vec{J}$ terms are sources of electromagnetic waves due to rotation of current density (vorticity of charged fluid elements) in the inhomogeneous electromagnetic wave equation, which is neglected in the MHD approximation because only slow phenomena are considered. The $\nabla \times \vec{J}$ terms show that kinetic helicity can be gained or lost by electromagnetic waves (as exploited by RF current drive) and is reflected in the species kinetic helicity by the first term of the right-hand side of Eq. (19).

Adding $m_\sigma \vec{u}_\sigma$ dotted with Faraday's equation and $\vec{B}$ dotted with Eq.(4), integrating with Leibniz' rule and multiplying by $q_\sigma$ gives the evolution of the species cross-helicity

$$m_\sigma q_\sigma \frac{d\mathcal{X}_\sigma}{dt} = 2q_\sigma^2 \int \vec{E} \cdot \vec{B} \, dV - 2q_\sigma \int \vec{B} \cdot \frac{\vec{R}_\sigma}{n_\sigma} \, dV - 2q_\sigma \int \vec{B} \cdot \frac{\nabla \mathcal{P}_\sigma}{n_\sigma} \, dV$$
$$- m_\sigma q_\sigma \int u_\sigma^2 \vec{B} \cdot d\vec{S} - 2m_\sigma q_\sigma \int \vec{\omega}_\sigma \cdot (\vec{u}_\sigma \times \vec{B}) dV - 2m_\sigma q_\sigma \int \vec{u}_\sigma \cdot \nabla \times \vec{E} \, dV \qquad (20)$$
$$+ 2m_\sigma q_\sigma \int (\vec{u}_\sigma \cdot \vec{B}) \vec{u}_\sigma \cdot d\vec{S}$$

and similarly, the magnetic helicity evolution multiplied by $q_\sigma^2$ is

$$q_\sigma^2 \frac{d\mathcal{K}}{dt} = -2 \, q_\sigma^2 \int \vec{E} \cdot \vec{B} \, dV - 2q_\sigma^2 \int \phi \vec{B} \cdot d\vec{S} - q_\sigma^2 \int \vec{A} \times \frac{\partial \vec{A}}{\partial t} \cdot d\vec{S} + q_\sigma^2 \int (\vec{A} \cdot \vec{B}) \vec{u}_\sigma \cdot d\vec{S} \qquad (21)$$

which can all be added up and compared to Eqs. (3) and (13). The three equations (19), (20) and (21) together with Eq. (3) are equivalent to an MHD approach with a parallel electric field in the magnetic helicity evolution Eq. (21) replaced instead by the complete Ohm's law. The primary advantage of the two-fluid approach is clarity in topological concepts and a broader regime of applicability (e.g. across skin depth scales, reconnection regions, electromagnetic phenomena).

The Lorentz force which appears in the second and fourth terms of the right-hand side of Eq. (19) cancels with the same terms in Eq. (20), and couples the species kinetic helicity to the species cross-helicity. In turn, the parallel electric field term $\vec{E} \cdot \vec{B}$ on the right-hand side of Eq. (21) couples to the same term in the cross-helicity Eq. (20). From an MHD point-of-view, magnetic helicity decays via the parallel electric field $\vec{E} \cdot \vec{B}$ inside the system volume, where the electric field is approximated by various forms of Ohm's law. Each term in Ohm's law then contributes to magnetic helicity decay (resistivity, pressure, etc). From a two-fluid point-of-view, the picture is more precise. Ignoring for now all the "AC injection" terms (all terms with $\partial/\partial t$) and the motion of boundaries ($\vec{u}_\sigma = 0$), the parallel electric field in Eq. (21) couples magnetic helicity to cross-helicity, which in turn is coupled to fluid kinetic helicity through the Lorentz force terms in Eqs. (19) and (20). The decay of magnetic helicity via the parallel electric field $\vec{E} \cdot \vec{B}$ is a source of cross-helicity, as is the kinetic energy $u_\sigma^2$ across any open magnetic flux surfaces. Note that magnetic helicity does not *directly* decay with friction, since there are no $\vec{R}_\sigma$ terms in Eq. (21), instead magnetic helicity is first converted to cross-helicity. Cross-helicity then decays in part with a frictional force parallel to the magnetic field $\vec{R}_\sigma \cdot \vec{B}$ (if friction is present), or a parallel pressure gradient $\nabla \mathcal{P}_\sigma \cdot \vec{B}$ (if present), and couples as a source to kinetic helicity through the Lorentz force terms. In turn, kinetic helicity decays through friction parallel with vorticity $\vec{R}_\sigma \cdot \vec{\omega}_\sigma$ and grows with enthalpy $h_\sigma$ across open vorticity flux surfaces. In effect, the decay of magnetic helicity is mediated by cross-helicity towards generation of helical flows, and the extent to which magnetic helicity decays is determined by the size parameter as shown in Section V. Establishing a net enthalpy (battery effect) or a net AC injection across vorticity or magnetic flux surfaces can then sustain helical flows and magnetic helicity against dissipation. The scenario corresponds to tokamak H-mode sustainment by NBI heating [22] and reverse-field pinch (RFP) flow generation by internal reconnection events [23].

Combining Eqs.(19), (20) and (21) together with Eq. (3) retrieves Eq. (13). Each term of Eq. (13) in the evolution of canonical helicity can now be interpreted directly in relation to the other helicities. The enthalpy battery effect $\int h_\sigma \vec{\Omega}_\sigma \cdot d\vec{S}$ is split into kinetic helicity injection by a battery effect on the $f$ vorticity flux tube component $\int h_\sigma \vec{\omega}_\sigma \cdot d\vec{S}$, and a battery effect on the $\psi$ magnetic flux tube component which injects magnetic helicity with an electrostatic potential $\int \phi \vec{B} \cdot d\vec{S}$ and injects cross-helicity with kinetic energy or pressure on the magnetic flux tube $\int (u_\sigma^2 + d\mathcal{P}_\sigma/n_\sigma)\vec{B} \cdot d\vec{S}$. The AC injection term $\int \vec{P}_\sigma \times \partial \vec{P}_\sigma/\partial t \cdot d\vec{S}$ includes four components: the inductive electromagnetic term $\int \vec{A} \times \partial \vec{A}/\partial t \cdot d\vec{S}$ which injects magnetic helicity, the kinetic term $\int \vec{u}_\sigma \times \partial \vec{u}_\sigma/\partial t \cdot d\vec{S}$ which injects kinetic helicity, the RF term $\int \vec{u}_\sigma \times \partial \vec{A}/\partial t \cdot d\vec{S}$ which injects kinetic helicity as used in RF current drive, an unsteady mechanical term $\int \vec{A} \times \partial \vec{u}_\sigma/\partial t \cdot d\vec{S}$ which injects kinetic helicity with applied forces that affects the electromagnetic component of the canonical momentum.

If there is no flow in the system, for fixed boundaries and zero mass electrons, the total canonical helicity becomes the ordinary magnetic helicity, so summing Eqs. (13) for each species gives

$$\frac{d\mathbb{K}}{dt} = -4e^2 \int \eta \vec{J} \cdot \vec{B}\, dV - 4e^2 \int \phi \vec{B} \cdot d\vec{S} - 2e^2 \int \vec{A} \times \frac{\partial \vec{A}}{\partial t} \cdot d\vec{S} = 2e^2 \frac{d\mathcal{K}}{dt} \qquad (22)$$

where friction was taken to be resistive only, $\vec{R}_e \simeq \vec{R}_{ei} = -ne\eta \vec{J}$, so viscosity and the thermal Nernst effect were ignored. Eq. (22) retrieves the basis of current drive by magnetic helicity injection [24, 25, 26] or, in non-linear form, the MHD dynamo effect in mean-field theory [27]. If there is steady flow in the system ($\vec{u}_i \neq 0, \partial \vec{u}_i/\partial t = 0$), again for fixed boundaries and zero mass electrons, and to maintain a constant total canonical helicity, the sum of Eqs. (13) for each species becomes

$$2m_i \int h_i \vec{\omega}_i \cdot d\vec{S} + m_i e \int \vec{u}_i \times \frac{\partial \vec{A}}{\partial t} \cdot d\vec{S} + m_i e \int u_i^2 \vec{B} \cdot d\vec{S} + 4e^2 \int \phi \vec{B} \cdot d\vec{S} + 2e^2 \int \vec{A} \times \frac{\partial \vec{A}}{\partial t} \cdot d\vec{S}$$
$$= -2m_i e \int \eta \vec{J} \cdot \vec{\omega}_i \, dV - 4e^2 \int \eta \vec{J} \cdot \vec{B} \, dV \qquad (23)$$

where enthalpy drive or AC injection applied to vorticity flow tubes or magnetic flux tubes balances the resistive dissipation in vorticity flow tubes or magnetic flux tubes. Viscous or thermal contributions would appear as extra sink terms on the right-hand side. In effect, Eq. (23) would form the basis for flow drive and current drive from enthalpy or inductive boundary conditions, and in non-linear form, presumably for a two-fluid dynamo effect (flow helicity converted into magnetic helicity) or zonal flows (magnetic helicity converted into flow helicity).

## VII. SUMMARY

The transport equations for relative canonical helicity developed in Sections II and III generalize previous formulations of isolated self-helicity to canonical flux tubes that can intercept system boundaries. Even if the system is isolated, it is still necessary to use gauge-invariant relative canonical helicity to examine the interaction between electron and ion fluids in the two-fluid framework. The flow field of one species must be considered as a reference field for the other species. Section IV showed that even if the system is isolated and frictionless, canonical helicity can be transferred from one fluid to another while the total canonical helicity remains approximately constant, provided non-equipotential enthalpy exists on a surface common to both ion and electron canonical flux tubes. This effectively couples electron and ion canonical helicity, generalizing previous treatments which argued for invariance of non-relative self-helicity, and therefore considered both as independent from each other. The two-fluid model then explains the dependence on the experimental size parameter of compact torus bifurcation (Section V). The slingshot effect during counter-helicity merging generates enthalpy at the surfaces of canonical flux tubes which then inject helicity into both magnetic and vorticity flow components. The final compact torus is an FRC if dominated by canonical helicity injection into the vorticity flow tube component and is a spheromak if dominated by canonical helicity injection into the magnetic flux tube component. The ratio between the two injection channels depends on the size parameter. The model provides a more fundamental constraint for the dynamics of plasma relaxation than simple magnetic helicity conservation, which is only relevant to quasi-static systems. The model retrieves all previous results for static plasmas and extends the same topological concepts to regimes where flowing two-fluid plasmas are applicable. Intuition about twists, writhes and links taken from magnetic flux tube evolution can be applied to canonical flux tubes and, in the reduced two-fluid regime, between magnetic flux tubes and vorticity flow tubes—in particular, reconnection of magnetic flux tubes to vorticity flow tubes. Finally, Section VI details the explicit coupling between magnetic, kinetic and cross-helicities to highlight how a form of helicity is transferred to other forms of helicity. Thus, the transport of relative canonical helicity constrains the interaction between plasma flows and magnetic fields and, in principle, should provide a more general framework for driving flows and currents from enthalpy or inductive boundary conditions.

**ACKNOWLEDGEMENTS**

The author acknowledges Y. Ono for sparking the initial thoughts on compact torus bifurcation and support from the Japan Society for the Promotion of Science (JSPS) during the initial part of this work.

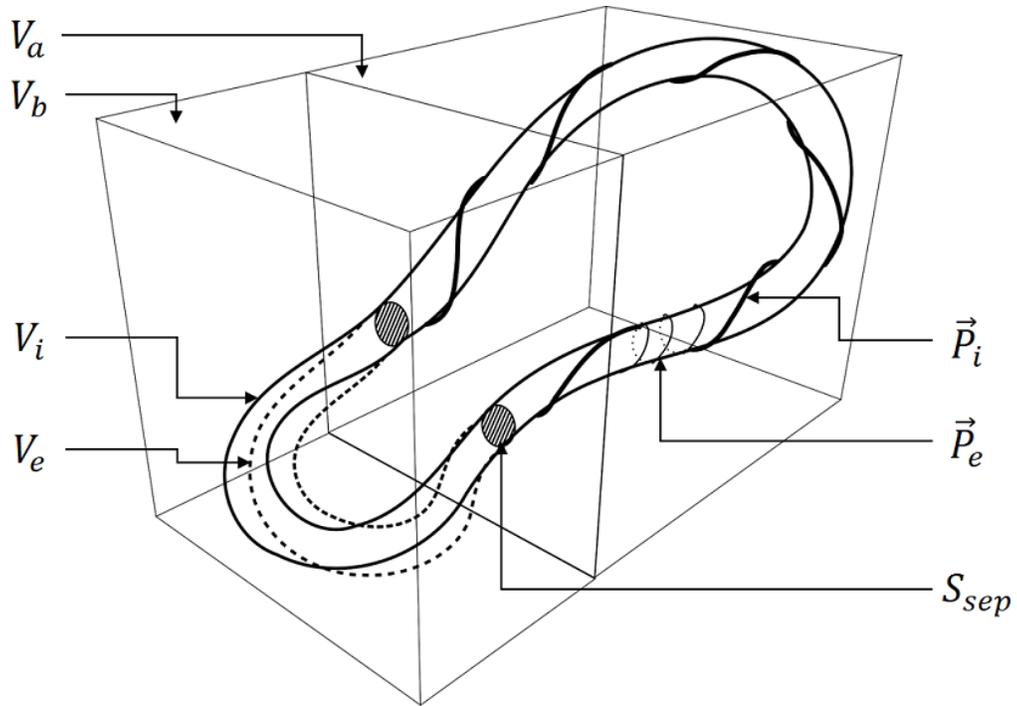

FIG. 1. Schematic of ion and electron canonical flux tubes $V_i$, $V_e$ defined by their canonical momentum $\vec{P}_i$, $\vec{P}_e$. The canonical flux tubes are topologically indistinguishable in volume $V_a$ (e.g. scales larger than the ion skin depth) but distinguishable in volume $V_b$ (e.g. scales less than the ion skin depth) and separated by a common surface $S_{sep}$. Relative canonical helicity must be used to compute helicity. An enthalpy difference across $S_{sep}$ injects canonical helicity into each flux tube at a rate determined by the size parameter $S^*$, see Eqs. (14), (17). If the whole system $V_a + V_b$ is isolated, then the electromagnetic component of canonical helicity does not require reference vacuum fields. The concept can be extended to scales below the electron skin depth where the electron canonical flux tube is topologically distinguishable from the magnetic flux tube.

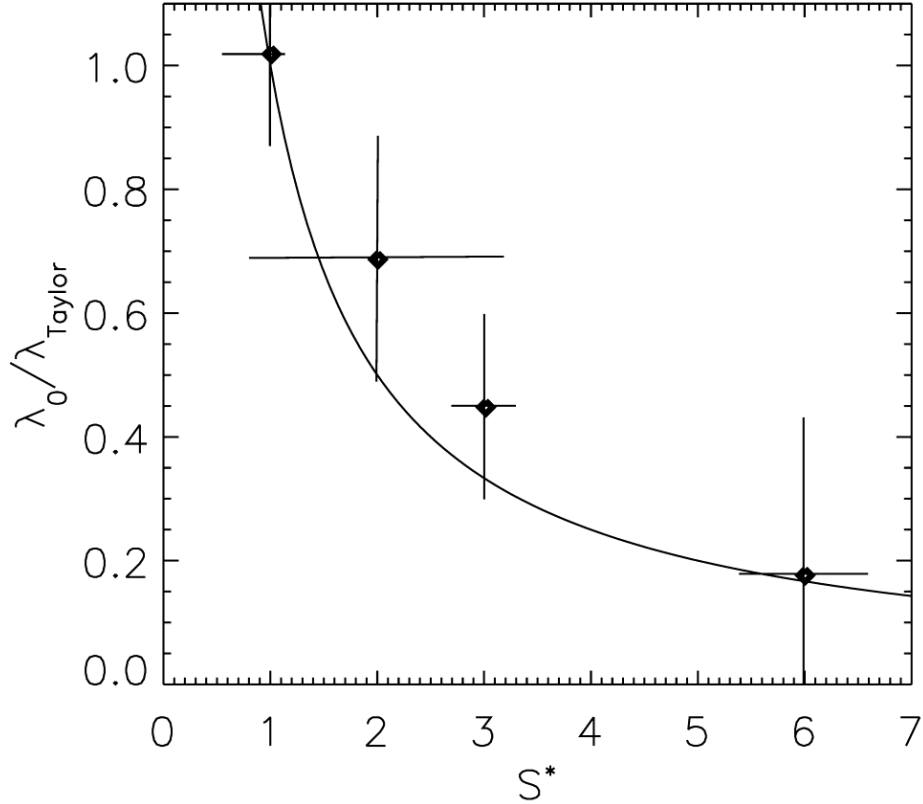

FIG. 2. Threshold initial poloidal eigenvalue $\lambda_0/\lambda_{Taylor}$ as a function of the size parameter $S^*$ for forming a spheromak or an FRC from counter-helicity merging. The threshold window for forming an FRC ($0 \leq \lambda_0/\lambda_{Taylor} \leq 1/S^*$) decreases as $S^*$ increases. The solid curve is given by Eq. (17) and explains the $1/S^*$ dependence with canonical helicity injection into the flow vorticity flux tube or the magnetic flux tube components of canonical flux tubes. Canonical helicity injection occurs due to enthalpy generation from the slingshot effect. Reprinted with permission from E. Kawamori, Y. Murata, K. Umeda, D. Hirota, T. Ogawa, T. Sumikawa, T. Iwama, K. Ishii, T. Kado, T. Itagaki, M. Katsurai, A. Balandin and Y. Ono, Nucl. Fusion 45, 843 (2005) Copyright (2005) IOP Publishing.